\documentclass[aps,superscriptaddress,showpacs,nofootinbib,eqsecnum]{revtex4}
\usepackage{amsmath,lscape,epsfig} \usepackage{amsfonts} 
\usepackage{amssymb} \usepackage{graphicx} \usepackage{multirow}
\def\be{\begin{equation}}
\def\ee{\end{equation}}
\def\bea{\begin{eqnarray}}
\def\eea{\end{eqnarray}}
\def\ms{\tiny \overline{\mbox{MS}}}
\def\MS{\overline{\mbox{MS}}}
\def\ds{\displaystyle}
\def\t{\tilde }
\def\ga0{\gamma_0}
\def\ga1{\gamma_1}
\def\ga2{\gamma_2}
\newcommand{\lsim}{\raisebox{-0.13cm}{~\shortstack{$<$ \\[-0.07cm] $\sim$}}~}
\newcommand{\gsim}{\raisebox{-0.13cm}{~\shortstack{$>$ \\[-0.07cm] $\sim$}}~}

\begin{document}
\baselineskip .5cm

\title{ Renormalization Group Improved Optimized Perturbation Theory:\\ Revisiting 
 the Mass Gap of the $O(2N)$ Gross-Neveu Model\\}

\author{Jean-Lo\"{\i}c Kneur}
\author{Andr\'e Neveu}
\affiliation{Laboratoire de Physique Th\'eorique et Astroparticules, CNRS, Universit\'{e} Montpellier 2, France}

\begin{abstract}
 We introduce an extension of a variationally optimized perturbation method, 
by combining it with renormalization group properties in a straightforward (perturbative) form. 
This leads to a very transparent and efficient procedure, with a clear improvement of the 
non-perturbative results with respect to previous similar
variational approaches. This is illustrated
here by deriving optimized results for the mass gap of the $O(2N)$ Gross-Neveu model, compared with
the exactly know results for arbitrary $N$. At large $N$, the exact result is reproduced already at the very first
order of the modified perturbation using this procedure. For arbitrary values of $N$, using the original
perturbative information only known at two-loop order, we obtain a controllable percent accuracy or less, 
for any $N$
value, as compared with the exactly known result for the mass gap from the thermodynamical Bethe Ansatz.  
The procedure is very general and can be extended straightforwardly to any renormalizable Lagrangian
model, being systematically improvable provided that a knowledge of enough perturbative
orders of the relevant quantities is available. 
\end{abstract}


\maketitle

\section{Introduction}
The variationally improved or optimized
perturbation (OPT) is by now a rather well-used modification of standard perturbation theory (for a
far from complete list of early references, see {\it e.g.}~\cite{delta}).  
It is based on a reorganization of the interacting Lagrangian such that it depends on
an arbitrary (mass) parameter (so-called linear $\delta$ expansion (LDE) in its simplest form), 
to be fixed by a definite optimization prescription, but it has
many other variants~\cite{kleinert1,odm,pms}.   
In $D=1$ theories, such as the quantum mechanical anharmonic
oscillator~\cite{ao}, described by a scalar $\phi^4$ field theory, the LDE turns out to be very similar~\cite{deltaconv}
to the ``order-dependent mapping" (ODM) resummation method~\cite{odm}, being equivalent at large orders 
to a rescaling of the adjustable oscillator
mass with perturbative order, which can suppress the
factorial large-order behaviour of ordinary perturbative coefficients. 
This appropriate rescaling of the adjustable mass 
gives a convergent series~\cite{deltaconv,deltac}
{\it e.g.} for the oscillator energy levels~\cite{ao} and related quantities. 
For the oscillator field theory, no renormalization
is needed, moreover the (ordinary) perturbative series is known to arbitrary high
orders, and the known large order behaviour of the series is a crucial guide 
both numerically and analytically to establish such convergence properties. 
In contrast, for most $D>1$ models, things become more involved due to the 
necessary renormalization, but the OPT
procedure can be made fully consistent~\cite{gn2} with the
renormalization program of ordinary perturbation theory, at least in a minimal subtraction scheme. 
Accordingly,  the prescription is well-defined on renormalized Lagrangians with 
appropriate counterterms. It is such that
any physical quantity whose ordinary renormalized perturbative series is available can then
be evaluated to order $\delta^k$ using well-defined modified Feynman rules.  
However in most $D>1$ renormalizable models
the perturbative series is only available for a few first orders, such that one can hardly study
convergence properties of the OPT method. 
Nevertheless it gives at least a well-defined systematically improvable 
way to go beyond mean field approximation and has a wide range of applicability, 
also at finite temperature and density.
To quote just a few examples of successful applications, for instance the occurence and precise location of a tricritical point and mixed
liquid-gas phase within the GN model~\cite{GN} in $D=2+1$ dimensions~\cite{prdgn3d} have been obtained, 
that was hinted at by Monte-Carlo simulations~\cite{gn3mc} but completely 
missed by the mean field approximation. OPT has also been used very recently~\cite{optNJL} 
in the non-renormalizable $D=4$ Nambu-Jona-Lasino model~\cite{NJL} to study the phase
transitions beyond mean field approximation 
in this simplified picture of low-energy QCD. In a different context, results   
for the shift of the critical temperature due to interactions in the Bose-Einstein condensate have been  
obtained~\cite{bec1,braaten,kleinert2,kastening1,kastening2,bec2}, with some results~\cite{kastening2,bec2} in remarkable
agreement with precise results from Monte-Carlo lattice simulations~\cite{mcbec}. In this latter case 
the relevant field model
is the three-dimensional super-renormalizable $O(N)$ $ \Phi^4$ model~\cite{zinnbook,arnold,baymN}, 
which somewhat simplifies renormalization issues, and the perturbative series is
known to high (seventh) order, which also allows
some definite conclusions on convergence properties\cite{kastening2,bec2}. \\

Now, in a more general renormalizable theory, 
once the renormalization procedure is well-defined, 
it is highly desirable to examine how the renormalization group (RG)
invariance of the full theory is realized within this optimized perturbation. More practically, 
it is also of interest to examine how
to incorporate eventually more information on higher orders from the RG within the calculated quantities. 
However in order to keep the RG resummation full information 
within the delta-expansion, it turns out that one has
to resum to {\em all} orders in the latter expansion. 
This can be done~\cite{gn1,gn2,qcd1,qcd2}, at the same level of approximation 
than RG is treated ({\it i.e.} typically  for
the leading, next-to-leading, etc\ldots logarithmic dependence on the mass parameter), but implying  
a rather involved formalism, rendering
the practical optimization with respect to the perturbative mass parameter not much   
straightforward (and only numerical). 
Numerical estimates in reasonable agreement~\cite{gn2} 
with the exactly known mass gap of the $O(2N)$ Gross-Neveu
(GN) model have been obtained in this RG-resummed $\delta$-expansion (at the few percent accuracy 
level). The method has also been applied to the QCD basic Lagrangian, and
results have been obtained for relevant chiral symmetry-breaking quantities
like the (constituant) quark mass, the quark chiral condensate and pion decay constant~\cite{qcd1,qcd2}, 
using only the (original) perturbative informations available at second order for those quantities. 
However, admittedly there is not much evidence for a systematic
control of the convergence and error of the method from those results, even {\it e.g.} in the GN  
case where a precise quantitative comparison
with known exact mass gap results for arbitrary $N$ values can be addressed. For instance
in the GN model, the discrepancy
with the exact mass gap $M(N)$ depended quite a lot on $N$, and also on the practical method used to 
optimize the mass
parameter\footnote{Typically most of our results in refs.~\cite{gn2,qcd1,qcd2}
depended on further using Pad\'e approximants for the mass dependence, representing rather specific ansatzes 
for the relevant chiral symmetric limit.}.\\
It was also shown~\cite{Bconv} that at large perturbative orders the method gives a
damping of the generic factorial growth of the large order perturbative coefficients (those
due to the usual renormalons), but this modification is not sufficient to establish formal convergence of
the new series,  
due to the extra complication brought about by logarithmic dependence in the mass in such renormalizable theories. 
A Borel resummation can improve the situation~\cite{Bconv}, but in any case    
such essentially qualitative large order results are  
of limited practical use to determine precise non-perturbative predictions for 
relevant physical quantities, for which in most cases only the very first few order
perturbative coefficients are known exactly.  \\

Despite the lack of a rigorous convergence proof of the method
for arbitrary renormalizable models, one may try to improve the basic method in order to
obtain eventually more efficient prescriptions and better approximations. In 
this paper we reconsider the basic LDE-OPT construction, but 
augmented from the very beginning by the information on RG invariance properties of the physical quantities
that are calculated. 
More precisely, we require the RG invariance to hold at the purely {\em perturbative} level of 
the $\delta$-expansion 
and in a most straightforward manner {\it i.e.} consistently at the same order 
as the perturbative (non-RG) available information. This is in a sense a less ambitious program
than constructing explicitly RG-resummed quantities, but as we will see it has the advantage of 
giving a very transparent procedure with RG properties `enrooted' at any step, 
and using nothing else than the available original purely perturbative information.
This new procedure has immediately as consequence 
to link the mass and coupling parameters, in a way similar to the standard RG behaviour. 
But in addition, since optimization also fixes the perturbative mass parameter, it implies that both  
the mass and the coupling are fixed, and there are no free parameters, for any given values
of the optimized perturbative mass. Of course, this is only due to fixing from optimization a
value of the variational mass, and the resulting fixed coupling is similarly a `variational'
parameter. But those variational mass and coupling are to be simply replaced in 
the relevant physical quantity being optimized, giving a 
 relation {\it e.g.} between the latter and the basic scale
$\Lambda_{\ms}$ of the model, typically given in the $\MS$ scheme. Actually, 
when such RG-improved OPT is applied to the dimensionless ratio of the mass gap to
$\Lambda_{\ms}$, it is completely
equivalent to optimizing independently with respect to the two mass and 
coupling parameters.  All those properties will become clearer as 
the procedure will be worked out below on a definite model.

Having thus summarized some previous developments of the ordinary method, 
the present work is organized as follows.  Sec. II briefly reviews some known perturbative and non-perturbative
results for the $O(2N)$ Gross-Neveu model and sets our conventions, defining also RG quantities to be used
later on. In Sec. III we present the LDE method and
the interpolated GN model, supplemented with the new idea on the incorporation of renormalization
group requirements.  For illustrative purposes we first define
the ingredient of the new RG improved variational method on the particularly simple $N \to \infty$ case, 
where both the exact mass gap is known~\cite{GN} and its perturbative expansion can be derived to arbitrary
large order. In this case the new procedure is
particularly simple and entirely analytic. It displays remarkable convergence properties as the order of the perturbation expansion increases. We discuss it in some detail as the generic 
behaviour of the solutions serves as a very useful guide to the more general case of arbitrary $N$.
In section IV we extend this framework to the case of arbitrary $N$ values, examining
also as an intermediate step the next-to-leading $1/N$ case. We determine approximations to the mass
gap for arbitrary values of $N$ and compare those with exact known results to estimate the error of
the method. We discuss some difficulties encountered and a few variants of the method. 
Finally conclusions are given in section V. 

\section{GN model perturbative and non-perturbative results} 
The $O(2N)$ Gross-Neveu model~\cite{GN} is described by the Lagrangian density for a fermion
field given by
\begin{equation} 
{\cal L} = 
\bar{\Psi} \left( i \not\!\partial\right) \Psi  
-m {\bar \Psi} \Psi 
+ \frac {1}{2} g^2_{GN}\: ({\bar \Psi} \Psi)^2 +{\cal L}_{C.T.}
\label{GN} 
\end{equation} 
where $\Psi$ is a $O(2N)$ vector and summation over $N$ is implicit in the above equation. 
In $1+1$ space-time dimensions the $\psi_k$ components of $\Psi$ for $k=1,..N$ represent two-component Majorana spinors. The mass $m$ and coupling $g^2_{GN}$ designate implicitely renormalized quantities
and the appropriate counterterm are not shown in Eq. (\ref{GN}). \\

When $m=0$, in addition to the $O(2N)$ symmetry  the theory
is invariant under the discrete chiral symmetry (CS)
\begin{equation} 
\Psi \to \gamma_5 \Psi \,\,\,, 
\end{equation} 
which is spontaneously broken such that the fermions get a non-zero mass~\cite{GN}.\\
For studying the model Eq. (\ref{GN}) in the large-$N$ limit it is
convenient to define the four-fermion interaction as $g^2_{GN} N/\pi = \lambda$. Since
$g^2_{GN}$ vanishes like $1/N$ we study the theory in the large-$N$ limit with
fixed $\lambda$~\cite{GN}. 
At leading order of the $1/N$-expansion, 
this mass gap is simply 
\be
M_{N\to\infty} = \Lambda_{\ms}
\ee  
where in this normalization 
\be
\ds \Lambda_{\ms}\equiv  \bar\mu\: e^{-\,\frac{1}{\lambda(\bar\mu)}}
\ee 
in terms of the renormalized coupling $\lambda(\mu)$ in the $\MS$ scheme at the renormalization scale
$\bar\mu$. \\
Next, the  exact expression of the mass gap 
for arbitrary $N \ge 2$ values has been calculated~\cite{TBAGN} from the Thermodynamic Bethe Ansatz (TBA). 
The result is
\be
\ds
\frac{M(N)_{\mbox{exact}}}{\Lambda_{\ms}} = \frac{(4\,e)^{\frac{1}{2(N-1)}}}{\Gamma[1-\frac{1}{2(N-1)}]}
\label{Mgapex}
\ee
and also useful is its $1/N$ expansion at next-to-leading (NLO) order
\be
\frac{M(N)}{\Lambda_{\ms}}(\mbox{\small NLO $1/N$}) = 1+\frac{1+2\ln 2 -\gamma_E}{2N}
\label{Mgap1N}
\ee
where $\gamma_E$ is the Euler-Mascheroni constant. \\

Independently of those exact results for the massless theory, one can consider in the massive GN $O(2N)$ model 
the perturbative expansion of the pole mass $M$ in terms of the running mass $m\equiv m(\bar \mu)$, which 
is known at present only up to two-loop order:
\be
M^{(2-loop)}_{pert} = m \left[1+g (c^{\ms}_1-\gamma_0 L)+g^2 \left(c^{\ms}_2 +\left(\gamma^2_0-\gamma^{\ms}_1 
-c^{\ms}_1(\gamma_0+b_0)\right)\: L
+\frac{\gamma_0}{2}\,(\gamma_0+b_0)\: L^2\right)\:\right]
\label{Mgn2}
\ee
where we take the normalization: 
$g\equiv g(\bar\mu)=g^2_{GN}/\pi$. 
For the sake of generality 
we have made explicit in Eq. (\ref{Mgn2}) the dependence upon the RG beta function $\beta(g)$ 
and anomalous mass dimension $\gamma_m(g)$ coefficients, 
which are known up to three-loop order in the $\MS$ scheme~\cite{gracey}.
(NB the `$\MS$' indices in Eq. (\ref{Mgn2}) indicate 
that the relevant coefficients are scheme dependent, here given in the
$\MS$ scheme).  
One defines as usual:
\be
\beta(g)= -b_0 g^2 -b_1 g^3 -b^{\ms}_2 g^4\;;
\ee
and 
\be
\gamma_m(g)= \gamma_0 g +\gamma^{\ms}_1 g^2 +\gamma^{\ms}_2  g^3\;;
\ee
and for the $O(2N)$ GN model one has the particular values in our normalization:

\be
b_0 = N-1\;;\;\;\;b_1= -\frac{b_0}{2}\;;\;\;\;b^{\ms}_2= -\frac{1}{16}(N-1)(2N-7)\;;
\label{betadef}
\ee

\be
\gamma_0 = N-\frac{1}{2}\;;\;\;\;\gamma^{\ms}_1= -\frac{\gamma_0}{4}\;;\;\;\;\gamma^{\ms}_2 = 
-\frac{1}{16}(N-\frac{1}{2})(4N-3)\;.
\label{gammadef}
\ee

Also in Eq. (\ref{Mgn2}) the finite non-RG perturbative terms are~\cite{gn2,gracey,gracey2}:
\be
c^{\ms}_1=0\;;\;\;\;\; c^{\ms}_2 = (N-\frac{1}{2})(\frac{\pi^2}{12}-\frac{3}{16})\; ;
\ee

The full RG operator reads in this normalization
\be
\mbox{RG}\equiv \mu\frac{d}{d\,\mu} =
\mu\frac{\partial}{\partial\mu}+\beta(g)\frac{\partial}{\partial g}-\gamma_m(g)\,m
 \frac{\partial}{\partial m} \; ,
\label{RG}
\ee
which should give zero at some perturbative order $g^k$ (up to terms of higher orders
${\cal O}(g^{k+1})$), when applied to a physical RG-invariant quantity like the pole mass above. 
Finally we also give for reference the expression of the basic scale $\Lambda_{\ms}$, relevant for
instance in Eq. (\ref{Mgapex}), and 
defined in this normalization as\cite{TBAGN,gn2}
\be
{\ds 
\Lambda_{\ms} =\bar\mu e^{-\frac{1}{(N-1)\,g}}\:\left(\frac{N-1}{2}\:\frac{g}{1-g/2}\right)^{\frac{1}{2(N-1)}}\;.
}
\label{Lamms}
\ee 
\section{The interpolated massive GN model }
\subsection{the LDE method including RG information}
Let us first examine the standard implementation of the LDE procedure
within the GN model, before supplementing this construction with the new ingredient from
RG information.  According to the usual LDE interpolation prescription~\cite {delta},  
from the {\it original} four fermion theory, Eq. (\ref{GN}), we define the deformed Lagrangian,

\begin{equation}
{\cal L}(m,g,\delta) =\bar{\Psi} \left( i \not\!\partial\right) \Psi  
-m (1-\delta) {\bar \Psi} \Psi 
+ \delta \frac {g^2_{GN}}{2} ({\bar \Psi} \Psi)^2 +{\cal L}_{CT,\delta}
\label{GNlde}
\end{equation}
so that the new perturbation parameter $\delta$ interpolates between a free
massive theory for $\delta=0$ and the original massless interacting theory for $\delta=1$.
In Eq. (\ref{GNlde}) 
the counterterm Lagrangian density, ${\cal L}_{CT,\delta}$, has the same
polynomial form as in the original theory, while the
coefficients are allowed to depend on $\delta$. 
In fact for any given ordinary perturbative expression
in terms of the original renormalized mass and coupling, in the $\MS$ scheme, the LDE procedure is implemented 
simply by the substitutions
\begin{equation}
m \to m (1-\delta)\;,\;\;\; g\to g\,\delta\;,
\label{subst1}
\end{equation}
into the perturbative series of any relevant physical quantity, and re-expanding the result
to order $k$ in the new expansion parameter $\delta$. Next, $\delta$ is set
to the value $\delta=1$, such as to recover the original (massless) theory
at infinite order,
while at any finite order $k$ there remains a
dependence on $m$ in the LDE of the physical quantity. Considering for the latter
typically the case of the pole mass, expanded at order $k$: $M^{(k)}(m,\delta=1)$,  
the standard, mostly used prescription is an optimization criterion 
(principle of minimal sensitivity (PMS)~\cite{pms}), 
requiring at each successive perturbative orders $k$: 
\be
\frac{\partial}{\partial\,m} M^{(k)}(m,g,\delta=1)\vert_{m\equiv \tilde m} \equiv 0
\label{OPT}
\ee
thus defining an optimal $k$-dependent mass $\tilde m$ value. 
Up to now we have described what the standard procedure is in most similar studies, up to
some variants in the prescriptions to fix the mass\footnote{For example some other studies use
a so-called `fastly apparent convergence' (FAC) prescription, requiring at order $k$ the order $k+1$ to
vanish. Though this prescription is often simpler (analytically) than optimization with respect to
the mass, it needs the knowledge of a priori higher orders so is even less practicable than
optimization in models where only few perturbative terms are known. Moreover the mass optimization 
is best suited for our additional requirement to incorporate the RG information as we shall see.}. 
The main new ingredient that we add
to this procedure, is to supplement the standard mass-fixing prescription (\ref{OPT})
with a direct perturbative RG information. Namely, at any given order $k$ of the new LDE expansion, 
we impose, in addition to the mass optimization, that the modified LDE series satisfies a 
standard RG equation at the appropriate perturbative
order, {\it i.e.}
\be
{\rm RG} \left(M^{(k)}(m,g,\delta=1)\right) =0
\label{RGopt}
\ee 
where the RG operator is defined in Eq. (\ref{RG}). Now, combining it with Eq. (\ref{OPT}) immediately
shows that the RG equation takes a simpler, reduced form:
\be
\left[\mu\frac{\partial}{\partial\mu}+\beta(g)\frac{\partial}{\partial g}\right]M^{(k)}(m,g,\delta=1)=0
\label{RGred}
\ee
Note also that applying both Eqs. (\ref{OPT}) and (\ref{RGred}) completely fixes $m\equiv \t m$ 
and $g\equiv \t g$, since one has two constraints for two parameters in this case. 
A further conceptual simplification occurs, when considering directly the required ratio
$M^{(k)}/\Lambda_{\ms}$: In fact, since $\Lambda_{\ms}(g)$ satisfies by definition
\be
\left[\mu\frac{\partial}{\partial\mu}+\beta(g)\frac{\partial}{\partial g}\right]\:\Lambda_{\ms} \equiv 0
\ee
consistently at a given perturbative order for $\beta(g)$, it is easy to 
show that Eq. (\ref{OPT}) and (\ref{RGred}), are completely equivalent to the following:
\be
\frac{\partial}{\partial\,m} \left( \frac{M^k(m,g,\delta=1)}{\Lambda_{\ms}} \right) =0\;;\;\;\;
\frac{\partial}{\partial\,g} \left( \frac{M^k(m,g,\delta=1)}{\Lambda_{\ms}} \right) =0\;;\;\;\;
\label{optrg}
\ee
{\it i.e.} the procedure is equivalent to optimizing independently 
with respect to the two parameters of the theory!
\subsection{Application to the  $N\to\infty$ case}
Let us examine now in some detail how this new RG+OPT prescription works in the leading $N\to\infty$ limit of
the GN model.
The $N\to \infty$ limit is particularly suited for analysis because one knows
the perturbative expansion of the original theory for $m\ne0$ to arbitrary perturbative order. The latter 
expansion can be cast into the simple form, after redefining $\lambda\equiv N g^2_{GN}/\pi$: 
\be
M(m,\lambda) = m \left(1+\lambda \ln \frac{M}{\mu} \right)^{-1}
\label{MLNpert}
\ee
where $m\equiv m(\mu)$ and $\lambda\equiv \lambda(\mu)$ are the renormalized mass and coupling
in the $\MS$ scheme.  The simplicity of the GN model in the leading large $N$ limit is illustrated in the
compact form of Eq. (\ref{MLNpert}), 
containing only the leading logarithmic dependence on $m$, {\it i.e.} 
there are no other next-to-leading logarithmic
or non-logarithmic corrections
to the relation between the mass gap $M$ and the running perturbative mass $m$. 
Moreover the correct result $M =\Lambda_{\ms}$ in the chiral symmetric limit $m\to 0$ 
can be retrieved directly from algebraic 
manipulation of Eq. (\ref{MLNpert}),  noting that the latter equation can be rewritten
as~\cite{gn2,Bconv} 
\be
\frac{M(m)}{ \Lambda_{\ms}}= \; e^{W(\hat m/\Lambda_{\ms})}\to 1\;\;\mbox{for}\; m\to 0\;,
\ee
where we used $\Lambda_{\ms}\equiv \mu e^{-1/\lambda}$, 
$W(x)$ is the Lambert implicit function defined as $W(x) e^{W(x)} = x$, 
$\hat m\equiv m/\lambda$ is the scale invariant
mass, and in the last limit we used the properties $W(x)\simeq x $ for $x\to 0$.  
(NB clearly the result $M/\Lambda_{\ms}=1$ had been obtained previously by direct calculation in  
the original massless theory $m=0$~\cite{GN}, but the
previous algebraic manipulation makes clear the link with the perturbative RG-resummed form Eq. (\ref{MLNpert})
for the massive theory). \\ 
Now assume for a while that we only know at some finite order the perturbatively expanded Eq. (\ref{MLNpert}), in terms of $m$ at a given order $\lambda^k$, and  
let us examine the results  of the modified LDE: after the substitution (\ref{subst1})
we obtain to first order in $\delta$:
\be
M^{(1)}(m,\lambda,\delta) = m \left(1-\delta (1+\lambda  \ln \frac{m}{\mu} )\right)\; .
\label{Md1LN}
\ee
Now taking $\delta\to 1$, the OPT equation (\ref{OPT}) gives:
\be
\lambda (1+ \ln \frac{m}{\mu})=0
\ee
with immediate solution (assuming $\lambda \ne 0$): 
\be 
\frac{\t m}{\mu} = e^{-1}\; .
\label{optLN}
\ee
At this stage, a standard treatment
of the LDE+PMS would consist in replacing this optimal mass value within the expression
of $M^{(1)}(m,\lambda,1)$ and to proceed similarly at successive orders of the expansion. 
This gives a result for the mass with a non-trivial dependence in $\lambda$ very different
from the perturbative one. For instance at first order it is simply:
\be
M^{(1)}(m,g,1) \to \lambda\; \t m = \lambda \mu e^{-1} \; .
\label{M1LN}
\ee
However this does not yet give the required relation for $M$ as only function of $\Lambda_{\ms}$. 
Moreover for finite $N$, or similarly in another more complicated model, 
it is even less obvious to extract from such result the relevant ratio $M/\Lambda_{\ms}$, due
to the fact that the OPT equation gives a generally
involved $g$-coupling dependence, which is not directly related with 
$\Lambda_{\ms}$.
Let us examine what our new simple prescription gives for the above $N\to \infty$ series:
The $\beta$ function for large $N$ is simply given as $\beta(\lambda)= -\lambda^2$ (with no
higher order corrections) so that 
the extra RG equation takes the form:
\be
\left[\mu\frac{\partial}{\partial\mu}-\lambda^2\frac{\partial}{\partial \lambda}\right] M^{(k)}(m,\lambda,\delta=1)=0 \; ,
\ee
which gives at first order when applied on  expression (\ref{Md1LN}):
\be
m\,\lambda\:(1+\lambda \ln\frac{m}{\mu})=0 \; ,
\ee
which for $m\ne0$ and $\lambda\ne0$ gives the non-trivial solution
\be
\t \lambda=(\ln\frac{\mu}{m})^{-1}
\label{gLN}
\ee
which reminds of the perturbative expression of the running coupling $\lambda$ for $\mu\gg m$
(the exact running coupling for $N\to \infty$ being given by Eq. (\ref{gLN}) but with $m\to\Lambda_{\ms}$).
Next combining Eq. (\ref{gLN}) and Eq. (\ref{optLN}) gives the result:
\be
\ln\frac{\t m}{\mu}= -1\;;\;\;\;\t \lambda=1\;,
\label{solLN}
\ee
which upon replacing this solution in the corresponding expression for the pole mass at this perturbative
first $\delta$-order, Eq. (\ref{M1LN})  gives:
\be
\frac{M^{(1)}(\t m,\t \lambda,\delta=1)}{\Lambda_{\ms}(\t \lambda)}= 1\;,
\ee 
{\it i.e.} we obtain the exact mass gap result already at the very first order. Eq. (\ref{solLN})
also implies $\t m =\Lambda_{\ms}$, so that the exact running coupling is also obtained. \\
One may proceed to successive
orders, and in fact at any arbitrary perturbative order $\delta$
we obtain solutions (\ref{solLN}) giving thus the exact mass gap result. This is quite satisfactory, but it
is instructive to examine further the behaviour of such solutions. Actually, 
a definite drawback of the optimization prescription is that it involves
minimization of a polynomial equation of order $k$ in the relevant
mass parameter $m$ at perturbative order $\delta^k$, supplemented now by the extra
RG equation. It is clear that more and
more solutions are to be considered when 
increasing the order and this non-uniqueness of the optimized solution 
may require extra choice criteria. 
In fact, up to order $k=3$ in the $N\to \infty$ case the (multi)-solutions are degenerate
and give uniquely the solution of Eq. (\ref{solLN}) giving the correct result for the mass
gap. For instance at order $k=2$ one finds
\be
(1+\lambda L)(1+L)=0
\ee  
from RG Eq. (\ref{RGred}), where we defined for convenience $L\equiv \ln m/\mu$,
and 
\be
(1+\lambda +\lambda\, L\,(3+L))=0  
\ee
from OPT Eq. (\ref{OPT}). Although the two equations are now intimately related
and are solved together within our approach, to make the link with the standard RG properties
it is convenient to consider the RG equation as determining the coupling $\lambda$ as function of the mass dependence
$L$, while the OPT equation finally determines the optimal mass.  
But note that both possible solutions of the RG equation,
the `standard' RG behaviour $\lambda =-L^{-1}$, or the other solution
$L=-1$, lead finally to the same solution of Eq. (\ref{solLN}) once substitued
into the OPT equation (excluding the trivial solutions $m=0$ or/and $\lambda=0$).
Now things become different starting at third order of the $\delta$-expansion, $k=3$,
where spurious solutions appear in one or both equations. More precisely the RG
equation gives
\be
(1 + \lambda L) (1 + 2 \lambda + \lambda L (5 + 2 L)) =0\;.
\ee
So, while the previous $\lambda = -L^{-1}$ is still a solution, having the standard
RG behaviour, 
an extra solution: 
\be
\lambda = -(2+5L+2L^2)^{-1}
\label{RGN3}
\ee     
 appears, having clearly the wrong RG-behaviour even for large $L$.
Injecting this very odd solution into the OPT equation gives
\be
(1+L)^3\;(7+2L)=0\;.
\ee
So even if the expected solution $L=-1$ reappears (and recovering again $\lambda=1$
from Eq. (\ref{RGN3})), an extra solution remains, that we consider spurious, giving
\be
L=-\frac{7}{2}\;,\;\;\lambda=-\frac{1}{9}\;,
\ee
which finally gives $M^{(3)}(\mbox{spurious})/\Lambda_{\ms} = -\frac{1}{324}\: {\rm e}^{-(25/2)}
\simeq -1.15\times 10^{-8}$ which is clearly unphysical. This solution may be rejected 
even if not knowing the right exact result, on the
basis that it does not have the correct perturbative RG
behaviour (moreover giving a negative coupling $\t g$). But more generally this 
illustrates that the higher order and non-linear
equations implied by the optimization procedure may lead to spurious
solutions, as we shall also encounter in the less trivial situation of finite $N$. Also, high order polynomial equations often have unstable solutions, {\it i.e.} small changes
in the coefficients of the higher orders may induce large variations of these solutions.  
In that case, qualitative considerations using the expected RG behaviour plus other criteria may be necessary
to remove such spurious solutions. 

In summary at the leading order of the $1/N$-expansion,  
we obtain for arbitrary high perturbative order $k$ the following properties:\\
-the RG equation factorizes to the form
\be
(1+\lambda L) f^{(k)}(\lambda,L)=0 \; ,
\ee
{\it i.e.} whatever the form of $f^{(k)}(\lambda,L)$, 
$\lambda=-(L)^{-1}$ is always a solution.\\
-Injecting the latter solution into the OPT equations gives
\be
(1+L)^k =0\;,
\ee  
{\it i.e.} giving $k$-multiplicity roots of the form Eq. (\ref{solLN}) thus giving the exact
result $M^{(k)}=\Lambda_{\ms}$. \\
Moreover, in fact injecting this RG `perturbative' behaviour solution directly into $M^{(k)}(m,\lambda,1)$ simply 
gives at any $k$-order:
\be
M^{(k)}(m,\lambda,1)= m
\ee
without any extra correction, and having not yet used at this stage the OPT equation determining
$m$. This coincidence here between the `pole' mass $M$ and the perturbative
mass $m$, meaning there are no perturbative corrections, is certainly 
a peculiar property of the large $N$ limit, but nevertheless a non-trivial result
specific from this construction, and not obvious from the original form of Eq. (\ref{MLNpert}). 
A similar situation occurs in the large-$N$ limit of the $O(N)$ oscillator where 
one of the solutions of the optimization procedure at each
order gives the exact answer for the ground state energy with a similar increase 
in flatness around that optimum~\cite{deltac}.\\
-Finally we observe that for any $k\ge2$ the second derivative with respect to the 
mass is zero at the extrema points, {\it i.e.}
the extrema becomes more and more flat, as intuitively expected.\\

\begin{figure}[htb]
\begin{center}
\mbox{\epsfig{figure=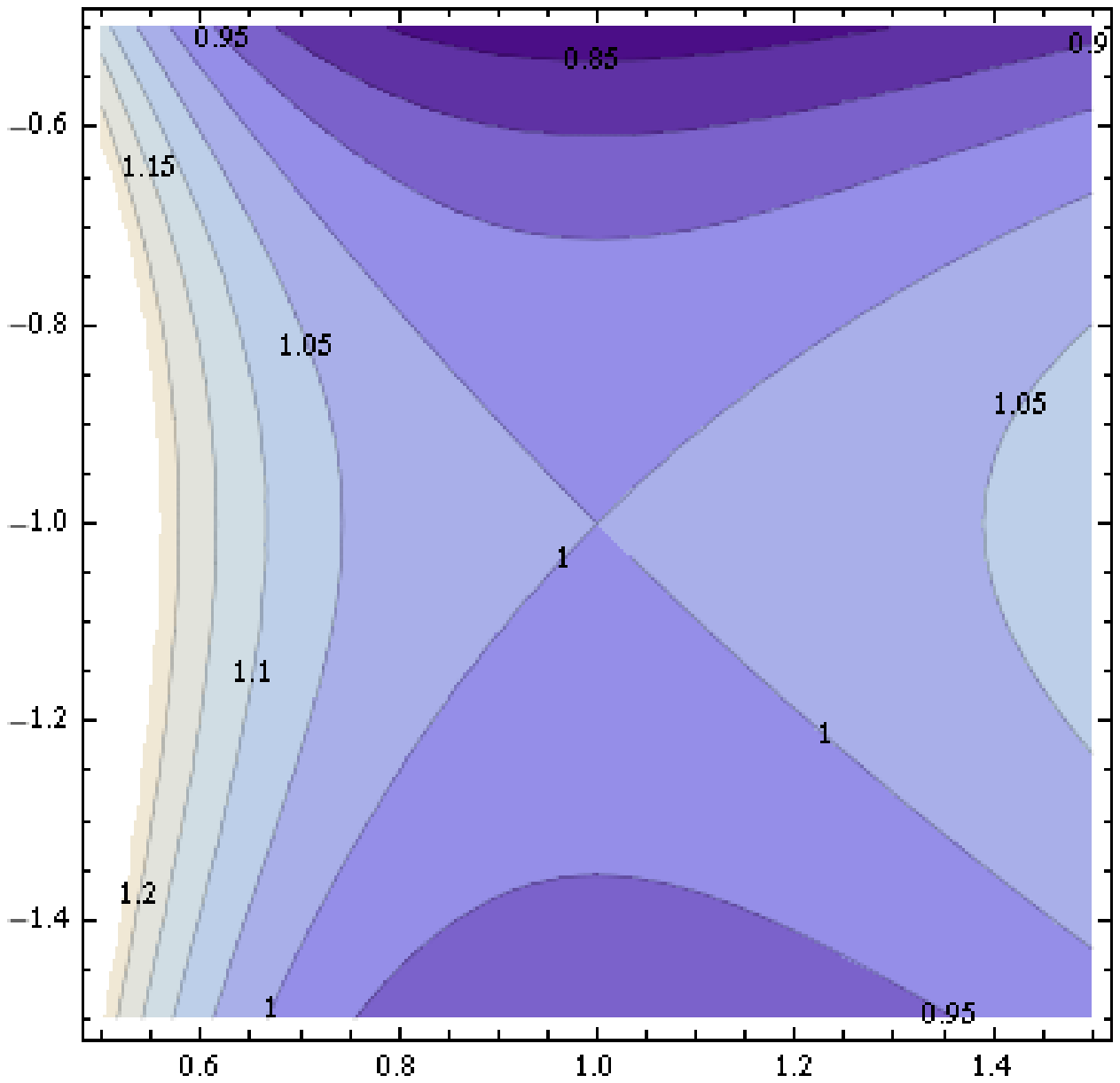,width=8cm}
\epsfig{figure=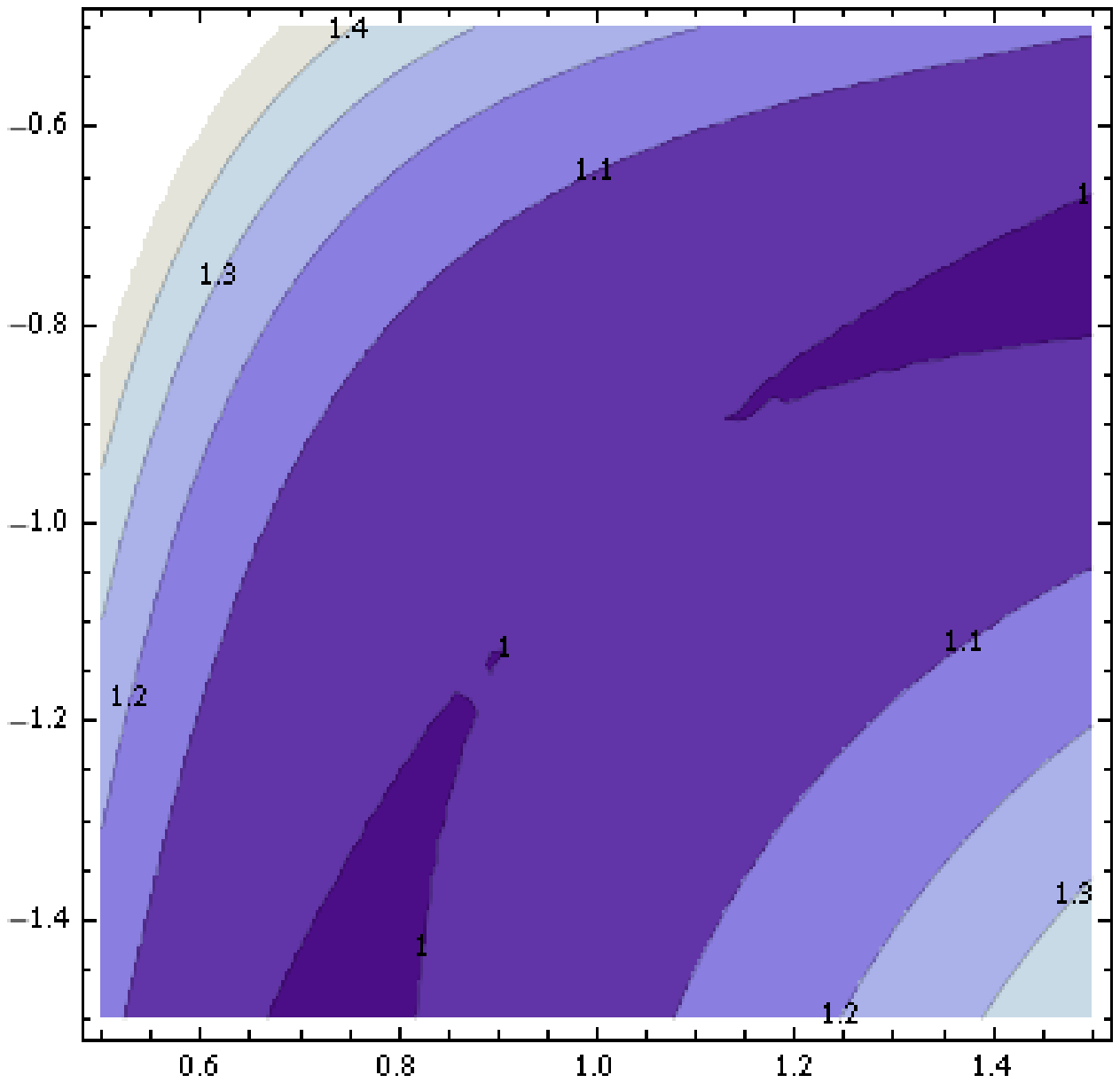,width=8cm}}
\vspace{-2cm}
\mbox{\epsfig{figure=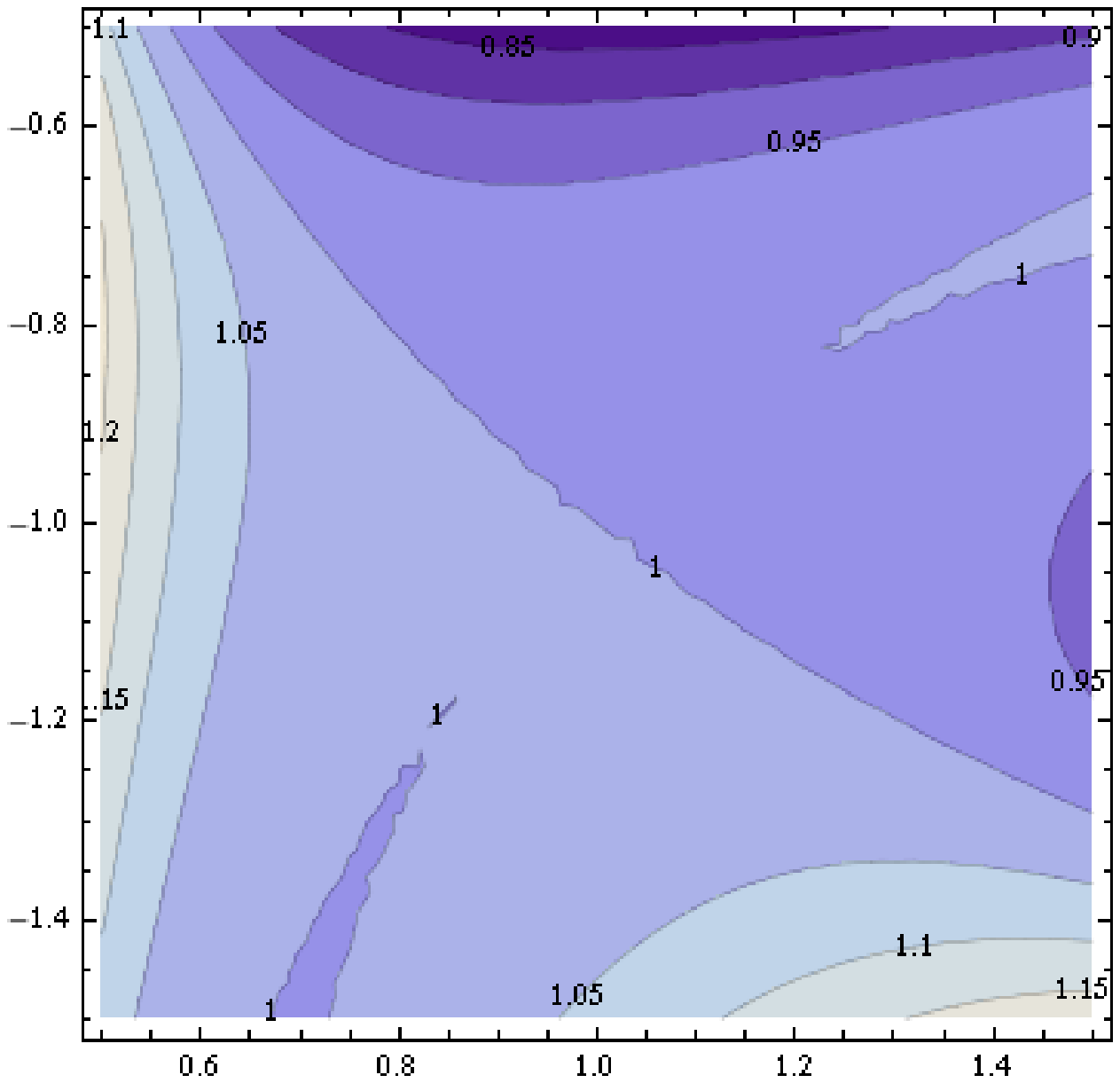,width=8cm}
\epsfig{figure=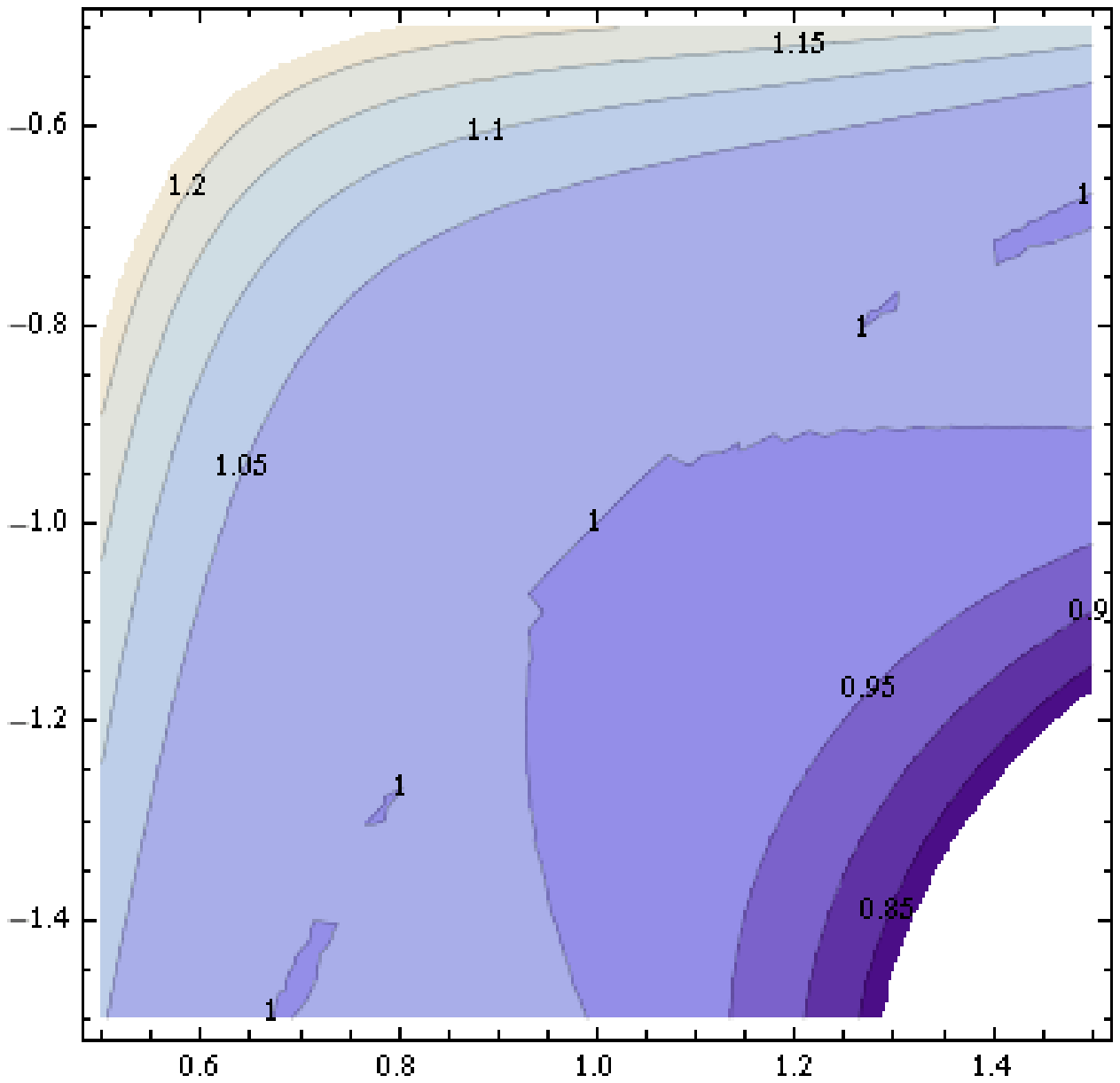,width=8cm}}
\caption{The surfaces $M^{(k)}$ as a function of $x=\lambda$ and $y=\ln\frac{m}{\mu}$
for $k=1\;, \;\;2\;, \;\;3\;, \;\;4$ (from top left to bottom right)}
\label{GNinf}
\end{center}
\end{figure}

To appreciate further the convergence properties of the procedure, we have plotted 
the surface $M^{(k)}(x\equiv\lambda,y\equiv \ln\frac{m}{\mu})$ for successive values of $k$ 
in Fig. \ref{GNinf} (using Mathematica~\cite{mathematica}). 
It is striking to see that this surface becomes flatter and flatter 
around the optimum $x=1, \; y=-1$. This is confirmed by the expansion of  $M^{(k)}$
around that point:

\bea
&M^{(1)}&= 1-\frac{1}{2}(x-y-2)(x+y)+ \mbox{ cubic terms}\; , \\ \nonumber
&M^{(2)}&= 1+\frac{1}{2}(x-y-2)^2+ \mbox{ cubic terms}\; , \\ \nonumber
&M^{(3)}&= 1-\frac{1}{6}(x-y-2)^2(4x+5y+1)+ \mbox{ quartic terms}\; , \\ \nonumber
&M^{(4)}&= 1-\frac{1}{2}(x-y-2)^3+ \mbox{ quartic terms}\; , \\ \nonumber
&M^{(5)}&= 1+\frac{5}{24}(x-y-2)^3(5x+7y+2)+ \mbox{quintic terms}\; ,\\ \nonumber
&M^{(9)}&=1+\frac{7}{16}(x-y-2)^5 (7x+11y+4) + \mbox{order 7 terms}\; ,\\ \nonumber
&M^{(10)}&=1+\frac{21}{16}(x-y-2)^6  +\mbox{ order 7 terms}\;.
\label{Ninfexp}
\eea

 At order $2k-1$ and $2k$ we may conjecture that there is around that point a behaviour of order 
$(k+1)$ in both $x$ and  $y$ directions,
corresponding to the increasingly flat behaviour which the figure suggests.   
We may see this property in the $N\to\infty$ case as an empirical convergence proof 
of the procedure for any value of $\lambda$ and $m$ in a large neighbourhood of 
$\lambda=1, \; \ln \frac{m}{\mu}=-1$.

All those properties of the large $N$ case will be a useful guide for the less trivial arbitrary
$N$ case.  
\section{Arbitrary $N$ case}
The modified LDE pertubation given from Eq. (\ref{subst1})
is now applied on the original
perturbative series given by (\ref{Mgn2}), only known exactly at two-loop order. 
In order to examine the eventual improvement and convergence properties of the method when increasing
the perturbative order, we shall compare the prescriptions at the available perturbative
orders, namely first and second
orders. We will also examine different prescriptions and approximations, also concerning the unknown higher orders
to estimate the sensitivity of our results to the latter. 
\subsection{OPT and RG at first order}
After substituting (\ref{subst1}) into Eq. (\ref{Mgn2}), expanding to first $\delta$-order
and taking $\delta=1$, one obtains
\be
\ds M^{(1)}(m,g,\delta=1)= -(N-\frac{1}{2})\,m\,g\, \ln\frac{m}{\bar\mu}\;.
\label{Mdel1}
\ee
We take the RG Eq. (\ref{RGred}) at first order for perturbative consistency, {\it i.e.}
only with the $b_0$ dependence entering. At this first order, both the OPT and RG Eqs. (\ref{OPT}), (\ref{RGred}), or equivalently
Eqs. (\ref{optrg}), respectively 
take a very simple form:
\be
g\:(L+1)=0
\label{opt1}
\ee
and
\be
m\,g\:(1+g L\,(N-1))=0
\label{rg1}
\ee
where $L\equiv \ln \frac{m}{\bar\mu}$. 
The unique non-trivial solution is: 
\be
\t L = -1\;;\hspace{1cm} \t g= -\frac{1}{(N-1)\,L} = \frac{1}{N-1}\;.
\label{rgd1}
\ee
Note that the solution of the OPT equation (\ref{opt1}) is similar to the $N \to \infty$ case above, 
while now the $N$-dependence enters through the RG solution (\ref{rg1}).  Moreover what is
quite remarkable is that just like in the $N\to\infty$ case, the RG solution again 
gives the correct RG behaviour of 
the running coupling for $\mu\gg m$: $g(\mu)=(b_0 \ln \frac{\mu}{m})^{-1}$, 
since Eq. (\ref{rgd1}) also implies $m=\Lambda^{(1)}_{\ms}$, 
noting that at first RG order, $\Lambda^{(1)}_{\ms}\equiv \bar\mu e^{-1/((N-1)\,g)}$. 
 Now, the correct comparison with the exact result Eq. (\ref{Mgapex}) implies
to use the very same normalization for $\Lambda_{\ms}$ entering Eq. (\ref{Mgapex}), 
defined in (\ref{Lamms}), and to substitute for $g$ the optimal solution (\ref{rgd1}): 
we thus obtain for the ratio of (\ref{Mdel1}) and $\Lambda_{\ms}$:
\be
\frac{M^{(1)}(\t m,\t g)}{\Lambda_{\ms}} = \frac{N-1/2}{N-1}\;\left(2 (1-\frac{1}{2\,(N-1)})
\right)^{\frac{1}{2\,(N-1)}}\;.
\label{M1dres}
\ee 
This gives $3/2, \sim1.38, \ldots$ for $N=2, 3, \ldots$ to be compared with the exact result
Eq. (\ref{Mgapex}), {\it e.g.} $\sim 1.8604, 1.4819,\ldots$ for $N=2, 3,\ldots$. Apart for
$N=2$ for which  it is a rather poor
approximation of the exact result, this is already quite reasonable for $N= 3$, 
and better for larger $N$ since the discrepancy decreases as $1/N$. (Note also that Eq. (\ref{M1dres}) is
zero if analytically continued to $N=3/2$, consistently with the exact result (\ref{Mgapex})).
More interestingly, the OPT first order result compares even much better with the next-to-leading order (NLO) 
in $1/N$ expansion of the mass gap: expanding to first $1/N$ order Eq. (\ref{M1dres}) 
one finds:
\be
\frac{M^{(1)}}{\Lambda_{\ms}}(\mbox{\small NLO $1/N$}) = 1+\frac{1+\ln 2}{2N}\;,
\label{Md1N}
\ee
to be compared with Eq. (\ref{Mgap1N}): the difference, $(\ln 2-\gamma_E)/(2N)\sim 0.058/N$ gives 
a relative error of Eq. (\ref{Md1N}) with respect to  
(\ref{Mgap1N}) of only $\sim 2\%$ for $N=2$ and much less for $N > 2$.
This result at NLO in $1/N$ can be attributed to the fact that the $b_0$ first order coefficient in
the RG equation is actually the only contribution, all other RG dependence  being ${\cal O}(1/N^2)$,
and thus the first $\delta$-order RG solution (\ref{rgd1}) turns out to give the correct complete RG dependence
at this $1/N$ order.

\subsection{$\delta$-expansion at second order }
We consider now the second order in the $\delta$-expansion applied to expression (\ref{Mgn2}) after substitution
(\ref{subst1}).    
At this order, the RG and OPT  
equations can still be managed analytically. Solving consistently at order $\delta^2$
we obtain respectively  for the 
OPT Eq. (\ref{OPT}):
\be
g\;\left[ 48 + g \left(-21 + 48 N + 12 L (-7 - 3 L + 4 (3 + L) N) + 4 \pi^2\right)\;\right] =0\;,
\label{opt2}
\ee
and for the RG equation (\ref{RGred}):
\bea
\label{rg2}
& m\:g^2\:\left[ 12 \left(5 + 6 L - 8 (1 + L) N \right) - 2 g (N-1) 
\left(-21 + 12 L (-1 - 3 L + 4 (1 + L) N) + 4 \pi^2\right) \right. \\ \nonumber
& \left.  +g^2(N-1) \left(-9 + 12 L (-1 - 3 L + 4 (1 + L) N) + 4 \pi^2\right)\:\right] =0\;.
\eea
(which again also exhibit trivial solutions $g=0$ and $m=0$ that we of course ignore). \\
Let us consider first the case $N=3$ as a typical moderate value ({\it i.e.} sufficently far from large $N$). 
Solving the two coupled RG and OPT equations (\ref{opt2}), (\ref{rg2}),  
one finds five different solutions: three give in fact complex-valued $\tilde m$, $\tilde g$ and 
$M^{(2)}(\tilde m,\tilde g)$, that we 
reject at the moment as {\it a priori} unwanted `spurious' solutions, and the two remaining real solutions are   
\bea
&g \simeq 1.69, L \simeq -2.52 \to \frac{M^{(2)}(\tilde m,\tilde g)}{\Lambda_{\ms}} \simeq 3.65\\ \nonumber
&g \simeq 0.41, L\simeq -1.74 \to \frac{M^{(2)}(\tilde m,\tilde g)}{\Lambda_{\ms}} \simeq 1.503
\eea
so that the second solution is very close, about $1.4\%$ above the exact result $\simeq 1.48185$.  
The other real solution is far away, and we now argue that it can be  
considered spurious. First, the relatively larger absolute values of $g$ and $L$ for this solution
may indicate that it comes from an excessive influence of the higher powers in the polynomial
 equations, which powers are not to be trusted at this order. Furthermore, a closer inspection of Eqs. 
(\ref{opt2}),(\ref{rg2}) is easy to do for any $N$, since
the RG equation is second order in $g$: expressing the two resulting solutions for $g(L)$, one realizes
that only one solution has the correct `perturbative' RG and large $N$ behaviour: expanding the solutions for
large $|L|$ gives:
\be
g^{(1)}(L)\simeq -\frac{1}{(N-1)L} +\frac{7N-6}{2(N-1)^2\,(4N-3)L^2} +{\cal O}(\frac{1}{L^3})
\label{g1sol}
\ee
where the first leading term exhibits the correct large $N$ and leading logarithm expression for the 
running coupling: $g(\mu)\sim (b_0 \ln \frac{\mu}{\t m})^{-1}$ for $\mu \gg \t m$ (provided that
$\t m/\Lambda_{\ms}$ remains of ${\cal O}(1)$, similarly to the leading and next-to-leading $1/N$
orders above), while the other solution gives
\be
g^{(2)}(L)\simeq 2 +\frac{1}{(N-1)L} +\frac{2+N-4N^2}{2(N-1)^2\,(4N-3)L^2} +{\cal O}(\frac{1}{L^3})
\label{g2sol}
\ee 
in clear conflict with both the known large $N$ behaviour $g\sim 1/N$ and a standard perturbative RG 
behaviour of the
coupling.  According to the lesson of the large $N$ limit, we are lead to select uniquely
the first solution, which indeed is the one giving $M/\Lambda_{\ms}(N=3) \simeq 1.503$. 
The same is observed for any $N\ge 3$ values: according to the RG behaviour criteria,
we uniquely select the first solution and this gives the result collected in Table \ref{tabN}.
We also give for illustration the corresponding values of $\t m/\Lambda_{\ms}$: one can remark 
that for arbitrary $N\ge 3$, the solutions are such that  
all $\t m/\Lambda_{\ms}$ values remains of order $1$ and regularly tends towards $1$ for 
increasing $N$, which is an a posterori crosscheck 
of the right perturbative RG behaviour of the solution (\ref{g1sol}). 
Moreover, the absolute
value of the second derivative with respect to the mass is smaller for this solution,
which is an additional indication of the expected behaviour similarly with the large $N$ above results. Also,  
in analogy with the $N\to\infty$ case, we can study for arbitrary $N$ the surfaces at second $\delta$-order,
$M^{(2)}(x\equiv g, y\equiv \ln m/\mu)$: let us just mention that 
from a direct minimization in the $(x, y)$ plane the same values for $N\ge 3$ of  
$M^{(2)}/\Lambda_{\ms}$ as those in Table \ref{tabN} are recovered, and that these surfaces exhibit an increasingly
flat behaviour around those extrema for increasing $N$. \\
%
\begin{table}
\begin{center}
\caption{\label{tabN} Combined OPT+RG results at second $\delta$-order, selecting solution (\ref{g1sol}), 
for the GN mass gap  for different $N$, with the corresponding
values of optimal mass parameter $\t m$ and coupling $\t g$, as compared with  
 $M^{exact}/\Lambda_{\ms}$.} 
\begin{tabular}{|c||c|c|c|c|c||c|}
\hline
$ N $ &$\frac{M^{(2)}(\t m,\t g)}{\Lambda_{\ms}} $ &  $ \t L\equiv \ln\frac{\t m}{\mu} $ & 
$\frac{\t m}{\Lambda_{\ms}}$ & $\t g$ &  $\frac{M_{exact}}{\Lambda_{\ms}} $ & \% error \\
\hline\hline
2   &   $ 1.69 \pm 0.09\,i$             &    $-2.28 \pm 0.18 i$   & &  $1.08 \pm 0.34 i$   &  1.86038 & \\
\hline
3   &    1.503                      & -1.74   & 0.70       & 0.41          &  1.48185 & 1.4 \% \\
\hline
4   &     1.338                        &    -1.57  & 0.81        &  0.27         & 1.3186  & 1.5\% \\
\hline
5  &     1.252                        &   -1.47 & 0.89      &   0.20            & 1.23668 & 1.2\% \\
\hline
8 &   1.143                         &   -1.33   & 0.92   &      0.12          & 1.1330 & 0.88\%\\
\hline
10 &      1.110           &                -1.28 & 0.95   &    0.094          & 1.10285 & 0.65\% \\
\hline
100 &   1.0099                      &    -1.019  & 1.005   &  0.0099         &  1.00916 & 0.07\%\\
\hline
\end{tabular}
\end{center}
\end{table}
A rather embarassing problem, however, occurs with lower values of $N$: actually, for $N=2$ we find only complex 
solutions when solving Eqs. (\ref{opt2},{\ref{rg2}). In table \ref{tabN} we put the result with the smallest
imaginary part, but even when taking the real part only it is not very close to the exact result, as compared
with $N\ge 3$ values obtained by the same procedure. We can examine this problem more closely by zooming
on $N$ values: since the TBA results are formally defined for any real $N>1$, we can always continue
this expression for non-integer $N$ values and compare to our approximation procedure, which is evidently
also well-defined for any real $N$. The results are given in Table \ref{tabNzoom} for a few representative $
2<N<3$ values. Our second-order approximation remains excellent for values $N<3$, until the point
when solutions becomes complex, which occurs at $N\simeq 2.1$ approximately. 
\begin{table}
\begin{center}
\caption{\label{tabNzoom} Same as in Table \ref{tabN}  for some representative real $2<N<3$ values.} 
\begin{tabular}{|c||c|c|c|c|c|}\hline
$ N $ &$\frac{M^{(2)}(\tilde m,\tilde g)}{\Lambda_{\ms}} $ &  $ \tilde L\equiv \ln\frac{\tilde m}{\mu} $ & $\tilde g$ 
&  $\frac{M_{exact}}{\Lambda_{\ms}} $ & \% error \\
\hline\hline
2.1   &    $1.72\pm 0.008 i$           &    $-2.26\pm 0.08i$       &  $1.08\pm 0.15i$      &  1.816 &  \\
\hline
2.2   &     1.800                        &    -2.37          &  1.31         & 1.768 & 1.8\% \\
\hline
2.4  &     1.679                        &       -1.96       &   0.642            & 1.677 & 0.1\% \\
\hline
2.6 &   1.614                         &   -1.87       &      0.537          & 1.600 & 0.9\% \\
\hline
2.8 &      1.554           &                -1.80     &    0.466          & 1.535 & 1.2\% \\
\hline
\end{tabular}
\end{center}
\end{table}
This reflects the fact that the optimization prescription Eq. (\ref{OPT}) alone already involves
a polynomial equation of order $k$ 
for $\ln m$, and the coupled OPT-RG equations (\ref{optrg}) become non-linear. As a result more and
more solutions, some of them being eventually complex, are to be considered when 
increasing the order, and  it is not much suprising that at second order the occurence of complex solutions
also depends on the value of $N$. 
In a previous work~\cite{bec2} we had proposed a rather simple way out for this problem with a
generalization of the PMS criterion as performed on the LDE series, which turns out to
lead to a drastic reduction of physically acceptable
real optimization solutions at each successive perturbative order. 
The modification is to introduce
extra variational parameters within the interpolating Lagrangian, starting here 
at the relevant order two with one more parameter $a$ introduced as 
\begin{equation}
m \to m\, (1 - \delta)^a \;.
\label{subst2}
\end{equation}
Next, the OPT criterion (\ref{OPT}) is generalized by requiring  
both the standard optimization (\ref{OPT}) {\em and}
\be
\partial^2 M^{(2)}/\partial m^2 =0
\label{OPTm2}
\ee
giving a system
of two equations to be solved simultaneously for $a$ and $m$, 
such that one can make real a pair of solutions which originally had a relatively small imaginary part for
the standard interpolation with $a=1$.  
At higher orders, the generalization is easily done with additional
parameters and additional vanishing of higher derivatives of 
$M(m)$. Such a modified prescription has been applied successfully {\it e.g.} to the 
calculation of the shift in critical temperature of the Bose-Einstein condensate (BEC) (where the knowledge of high 
perturbative orders
leads to a similar problem of complex OPT solutions), with excellent agreement
with lattice results~\cite{bec2}.\\
There are of course other possible perturbatively equivalent ways to 
introduce such an extra variational parameter, as long as the only 
constraint is that the modified Lagrangian
still interpolates between the free field (massive) theory for
$\delta=0$ and the original (massless) theory for $\delta=1$. The non-linear $\delta$-expansion with a 
simple exponent form in Eq. (\ref{subst2}) 
gives algebraically simpler subsequent optimisation and RG equations (\ref{optrg}), and is inspired from 
a prescription~\cite{bec2} suited to the $D=3$ $O(N)$ $g\Phi^4$ model case, where it is 
closely connected to other prescriptions, 
introducing explicitly the critical exponent related to the anomalous mass dimension within a 
power-modified interpolating mass~\cite{kleinert1,kleinert2,
kastening2}, which was argued to drastically improve the OPT convergence.  
(NB indeed the exact value of this critical exponent for the $D=3$ $O(N)$ $\Phi^4$ model 
at the next-to-leading $1/N$ order had been obtained empirically in \cite{bec2} by solving the analog of
Eq. (\ref{OPTm2}) together with the analog of the OPT Eq. (\ref{OPT}).} 
In the present $D=2$ GN case however, this modified OPT prescription is to be 
viewed as an essentially algebraic trick
to force one of the solutions to be real for $N\sim 2$, with a priori no particular physical meaning of 
the extra variational parameter $a$).

Using Eq. (\ref{subst2}) instead of (\ref{subst1}), 
 expanding  (\ref{Mgn2}) at order $\delta^2$, and applying
Eqs. (\ref{optrg}) and (\ref{OPTm2}) to solve for $a$, $m$ and $g$ gives the result in Table \ref{tabNa}
 for some representative values of $2<N<3$. We recover in this way real results for $N\sim 2$ 
which are very good for $N \lsim 2.5$. However, when $N$ is increasing, results
are somewhat worse than the original ones in Table \ref{tabNzoom} from the simplest OPT prescription. 
Of course, as long as
we directly obtain from the simplest prescription real solutions for any $N \gsim 2.1$ values, 
it is not needed to appeal to the more elaborate
interpolation with an extra parameter $a$: thus we obtain in this way an overall controllable 
prescription for any $N$.
Moreover even if we would not know the exact solution, it is easily seen in the present case that
the solutions for $a\ne 1$ progressively depart from the expected $1/N$ and perturbative RG behaviour
for increasing $N$ values.\\

\begin{table}
\begin{center}
\caption{\label{tabNa} Combined extended OPT+RG results at second $\delta$-order for the 
 GN mass gap  for some representative real $2<N<3$ values,
with corresponding values of the optimal parameters $a, \t m, \t g$, and compared with  
 $M^{exact}/\Lambda_{\ms}$.} 
\begin{tabular}{|c||c|c|c|c|c|c|}
\hline
$ N $ &$\frac{M^{(2)}(\tilde m,\tilde g)}{\Lambda_{\ms}} $ & a& $ \tilde L\equiv \ln\frac{\tilde m}{\mu} $ &
 $\tilde g$ &  $\frac{M_{exact}}{\Lambda_{\ms}} $ & \% error \\
\hline\hline
2 &     1.907 &   1.986  & -3.815  & 0.187  & 1.86038  &  +2.5\% \\
\hline
2.1   &  1.812   &  1.931 & -3.716 &  0.170    &   1.816 &  -0.2\%  \\
\hline
2.2   &     1.734  & 1.886 & -3.637  & 0.155   & 1.768 & -2\% \\
\hline
2.4  &     1.614   & 1.818  & -3.516  &   0.133  & 1.677 &  -3.9\% \\
\hline
2.6 &   1.526   &  1.77   &   -3.428  &   0.116      & 1.600 & -4.8\% \\
\hline
2.8 &      1.458   & 1.734   & -3.36  &    0.103       & 1.535 & -5.3\% \\
\hline
\end{tabular}
\end{center}
\end{table}

\subsection{Approximate schemes at second order}
In the previous second $\delta$-order calculation we required the exact two-loop order RG equation
(\ref{rg2}) to hold, which results in a second order equation in $g$ (after eliminating the trivial solution
$g=0$), thus responsible for the occurence of two solutions in Eqs.  (\ref{g1sol}),(\ref{g2sol}), one of those
being spurious according to the RG behaviour, or complex for small $N$ values. 
In fact one can think of two alternative ways to avoid this problem from the beginning, while still using the
RG perturbative information, noting that the purely perturbative
information at this available second order does not require the RG equation to hold exactly. 
Firstly, 
standard RG invariance only requires that the full RG operator (\ref{RG})
when applied to (\ref{Mgn2}), gives a remnant term of ${\cal O}(g^3)$, rather than being exactly zero,
since those higher order terms would be compensated by higher order (3-loop) terms in the pole mass itself,
which are not available. We can impose a similar requirement in the case of the $\delta$-expanded
pole mass expression, truncating perturbatively the reduced RG Eq. (\ref{RGred}) instead of requiring
it to hold exactly. 
Secondly, quite similarly, the expansion of the exact second order solution of
$g(L)$ for large $|L|$ in Eq. (\ref{g1sol}) formally gives an infinite series for the leading, next-to-leading, etc\ldots
logarithm $L$ dependence (LL, NLL). But, the genuine NNLL etc\ldots orders also need the knowledge of higher
(3-loop and beyond) RG behaviour, so that in ignorance of the latter, we may truncate Eq. (\ref{g1sol})
at the NLL Level.  Those two possible approximation schemes for the RG dependence are
not numerically equivalent, as we shall see, and such variants  
may also provide indirectly  an estimate of a theoretical error of the method at this second order. 
\subsubsection{Perturbative truncation of the RG equation}
In analogy with the ordinary RG properties, we only require a third-order truncated reduced RG Eq. (\ref{RGred}) 
(or equivalently Eq. (\ref{optrg}) for the RG part), when applied to the
variational $\delta$-expanded pole mass, since the fourth order actually contains RG terms 
that would cancel with three-loop terms in the pole mass. This third order truncation of Eq. (\ref{rg2})      
nevertheless keeps a dependence on the two-loop RG coefficients $b_1$ and $\gamma_1$, but now gives
a simple linear equation (omitting of course the trivial solution $g=0$) thus with a unique solution:
\be
\frac{5}{8}-N +(\frac{3}{4}-N)\,L -g\:\frac{(N-1)}{48}\left[ 21-4\pi^2 +12L\,(1 + 3 L - 4 (1 + L) N)\right]\;=0
\ee
which moreover is easily checked to have the correct large $N$ and RG perturbative behaviour: $\t g\sim -1/((N-1)L)$ for
large $|L|$. Combining this solution with the OPT equation gives a third order equation for $L$,
with a unique real solution,  which moreover is such that $\t m/\Lambda_{\ms}$ is very close to $1$
for any $N\ge 2$ values. This gives the results for arbitrary $N$ given 
in Table \ref{tabrgtrunc3}. We note that those are the closest results to the exact mass gap, with
the relative error being less than $1\%$ for any $N\ge 2$, and often well below the percent level. 
\begin{table}
\begin{center}
\caption{\label{tabrgtrunc3}Combined OPT+ third-order truncated RG at second $\delta$-order for the 
GN mass gap  for different $N$, with the corresponding
values of optimal mass parameter $\t m$ and coupling $\t g$, as compared with  
 $M^{exact}/\Lambda_{\ms}$.} 
\begin{tabular}{|c||c|c|c|c|c||c|}
\hline
$ N $ &$\frac{M^{(2)}(\t m,\t g)}{\Lambda_{\ms}} $ &  $ \t L\equiv \ln\frac{\t m}{\mu} $ & 
 $\frac{\t m}{\Lambda_{\ms}}$ & $\t g$ &  $\frac{M_{exact}}{\Lambda_{\ms}} $ & \% error \\
\hline\hline
2 & 1.84478        & -1.548  & 0.834 &     0.834 &  1.86038     & -0.86\%  \\
3 & 1.48460        &  -1.313  & 0.958 & 0.443   &   1.48185     & +0.19 \% \\
4 & 1.32627          &  -1.209 &  0.978 & 0.307  &    1.3186     &+0.58\%  \\
5 & 1.24448           & -1.138  & 0.984 & 0.239  &   1.23668     & +0.63 \%  \\
8 & 1.13956        & -0.962   &  0.979 & 0.158  &   1.133      &+0.58 \%  \\
10 & 1.10896        &  -0.952   & 0.981 & 0.123  &  1.10285      &+0.55 \% \\
100 & 1.00993       &  -0.958  & 0.998 & 0.0106  & 1.00916      &+0.077 \% \\
\hline 
\end{tabular}
\end{center}
\end{table}

\subsubsection{Truncation at the next-to-leading logarithm order approximation}
Next we examine here the results of a different kind of truncation, as discussed above, namely truncating
the right behaviour solution Eq (\ref{g1sol}) for large $L$ at the NLL level, invoking that 
in a standard perturbative framework, higher NNLL etc\ldots
orders need higher (three-loop) order information that we do not take into account in Eq. (\ref{Mgn2}). 
Combining this with the OPT Eq. (\ref{opt2}) gives the result in Table \ref{tabnlltrunc}. 
Those results are reasonably good, at the percent or so level, but not as good as the previous ones
obtained from truncating directly the RG equation.   
\begin{table}
\begin{center}
\caption{\label{tabnlltrunc}Combined OPT+ $g(L)$ dependence truncated at NLL level, at second $\delta$-order for the 
GN mass gap  for different $N$, with the corresponding
values of optimal mass parameter $\t m$ and coupling $\t g$, as compared with  
 $M^{exact}/\Lambda_{\ms}$.} 
\begin{tabular}{|c||c|c|c|c|c||c|}
\hline
$ N $ &$\frac{M^{(2)}(\t m,\t g)}{\Lambda_{\ms}} $ &  $ \t L\equiv \ln\frac{\t m}{\mu} $ & 
$\frac{\t m}{\Lambda_{\ms}}$  & $\t g$ &  $\frac{M_{exact}}{\Lambda_{\ms}} $ & \% error \\
\hline\hline
2 & 1.8663        & -1.77 & 0.69 &  0.82   &  1.86038     & +0.3\%  \\ 
3 & 1.503        &  -1.55  & 0.85 &  0.41  &   1.48185     & +1.4 \%  \\
4 & 1.337          &  -1.45 & 0.92 & 0.27  &    1.3186     &+1.4\%   \\
5 & 1.252           &  -1.39 & 0.90 &  0.21  &   1.23668     & +1.2 \%  \\
8 & 1.143        &  -1.29  & 0.96 & 0.12   &   1.133      &+0.9 \%   \\
10 & 1.1106        &   -1.26 & 0.95 & 0.095   &  1.10285      &+0.7 \%   \\
100 &  1.0099     &  -1.07 & 0.997 &  0.0095  &    1.00916   & +0.07\%  \\
\hline 
\end{tabular}
\end{center}
\end{table}

\subsection {Higher order estimates and stability}
Our results above give a clear evidence of numerically fast convergence of this RG-improved implementation
of the OPT method. 
In the absence of a rigorous convergence proof, we can still 
try to examine higher order estimates and the sensitivity and stability properties
of our results with respect to higher orders. The known three-loop RG beta function and anomalous
mass dimension in Eqs. (\ref{betadef},(\ref{gammadef}) may be used, giving all the logarithmic
dependence at order $g^3$ in an extension of Eq. (\ref{Mgn2}):
\bea
\label{Mgn3}
M^{(3-loop)}_{pert} = &m \left[ 1+g (c_1-\gamma_0 L)+g^2 \left(c^{\ms}_2 +\left(\gamma^2_0-\gamma^{\ms}_1 
-c_1(\gamma_0+b_0)\right)\: L
+\frac{\gamma_0}{2}\,(\gamma_0+b_0)\: L^2\right)\:\right. \\ \nonumber
 & \left. +g^3\left( a_{30}\,L^3 +a_{31}\,L^2+a_{32}\,L+c_3 \right)\:\right]
\eea
where the general expressions of the LL, NLL, and NNLL terms can be derived by applying RG properties:
\be
a_{30} = -\gamma_0/6 (\gamma_0+b_0)(\gamma_0+2b_0)
\ee
\be
a_{31} = c_1 (\gamma_0+b_0)(\gamma_0+2b_0)/2 +\gamma_1 (\gamma_0+b_0)+
\gamma_0(b_1-3/2 b_0 \gamma_0-\gamma_0^2)
\ee
\be
a_{32} = -c_2(\gamma_0+2b_0) -c_1 (b_1+\gamma_1-\gamma_0(\gamma_0+b_0)) -
\gamma_0^3 +2\gamma_0 \gamma_1-\gamma_2
\ee
with in the GN model $c_1=0$ and $c_2$ and the $b_i$ and $\gamma_i$ coefficients are given above. 
But the three-loop non-RG coefficient 
$c_3$ in Eq. (\ref{Mgn3}) is presently
unknown. To have a very conservative estimate of the latter, we do not even assume it to be positive,
and vary it between $0 < |c_3| < |c_3|_{max}$ where for $|c_3|_{max}$ we take 
 a ``theoretically-inspired'' maximal value, namely the coefficient given by the leading 
renormalon~\cite{renormalons} factorial behaviour at large perturbative order. This factorially growing 
behaviour of the purely perturbative GN pole mass has 
been estimated from graphs at the next-to-leading $1/N$ order in ref.~\cite{KRcancel}, and is
very similar to the well-known leading renormalon of the QCD pole mass~\cite{renormalons}.  
In the present normalization, it gives for the coefficient at perturbative order
$k$:
\be
|c_k| \sim \frac{1}{2} (k-1)! \:b_0^{k-1}  \;.
\label{leadrenorm}
\ee
Taking three representative extreme values, $c_3=0$ and from Eq. (\ref{leadrenorm}) 
$c_3 = \pm b^2_0 =\pm (N-1)^2$
as a crude estimate for the 3-loop coefficient $c_3$ in Eq. (\ref{Mgn3}), we applied the OPT and RG equations
on the third order $\delta$-expansion of Eq. (\ref{Mgn3}). This gives 
an OPT equation cubic in $L$ and a RG equation quartic
in $g$, thus as expected the combined higher-order equations (\ref{optrg}) have many solutions, 
most of those being complex and some solutions being quite unstable. But similary to the previous
case at order 2, it is not difficult to select
the much fewer solutions having the correct perturbative and large-$N$ behaviour,
with the results given in Table \ref{tabhe}.
For those well-behaved solutions, the complex results have reasonably small imaginary parts, 
but in this case we did not attempt to refine this analysis to 
recover real solutions with the above alternative prescriptions, since it is
anyway based on a crude estimate of an actually unknown coefficient. \\
%
\begin{table}
\begin{center}
\caption{\label{tabhe}Third order estimates for combined OPT+ RG  for the 
GN mass gap  for different $N$,  as compared with  
 $M^{exact}/\Lambda_{\ms}$, with  corresponding
values of the unknown three-loop non-RG coefficient $c_3$.} 
\begin{tabular}{|c||c|c|c||c|}
\hline
$N$  &$\frac{M^{(3)}(\t m,\t g)}{\Lambda_{\ms}} $ &$\frac{M^{(3)}(\t m,\t g)}{\Lambda_{\ms}} $ 
&$\frac{M^{(3)}(\t m,\t g)}{\Lambda_{\ms}} $  &  $\frac{M_{exact}}{\Lambda_{\ms}} $  \\
     & ($c_3=-b^2_0$)  &           ($c_3= 0 $)  & ($c_3 = b^2_0$) &   \\ 
\hline\hline
2 & 1.845 & 1.898        &    $1.96 \pm 0.03 i$   &  1.86038    \\
3 & 1.411 & $1.4666\pm 0.0005 i$     & 1.54 $\pm 0.04 i$      &   1.48185     \\
5 & 1.192 & $1.2388 \pm 0.008 i$     & 1.27$\pm0.04 i$     &  1.23668    \\
100 & 1.006 & $ 1.0108 \pm 0.001 i$  & 1.014 $\pm 0.004 i$    & 1.00916 \\ 
\hline 
\end{tabular}
\end{center}
\end{table}
%
Overall these results can be considered very stable:
a large variation of the third-order coefficient, $ -b^2_0 <c_3 <  b^2_0$,  
leads to very reasonable changes in the final result. Moreover even for the
extreme lower and upper bounds $\pm |c_3|_{max}$, the results bound very well
(and remain rather close to) the exact mass gap.
Note that the renormalon-inspired coefficient $|c_3|_{max}$ grows fastly with $N$, since $b_0 ={\cal O}(N)$,
but this does not prevent the mass gap result to be close to the exact one for sufficiently
large $N$. While some among the multi solutions of Eqs. (\ref{optrg}) are clearly unphysical or 
can reflect instabilities, the reason is that selecting the OPT +RG solution giving the correct large $N$ 
behaviour for the coupling, namely $g \sim 1/N$, 
implies that the $g^3\sim 1/N^3$ compensates for the renormalon $|c_3|\sim N^2$ behaviour for large $N$.  
We also checked finally that there are intermediate values of $ |c_3| < b^2_0$ such that
a well-behaved solution is real and very close to the exact mass gap (for instance this happens
for $c_3\simeq 0.61$ for $N=2$).\\
\section{Conclusions and outlook}
We have examined a very simple additional ingredient to the standard variational OPT approach,
in order to supplement 
it with RG perturbative information. The procedure is very transparent and largely analytical
at second order. It has as immediate consequence to fix both  
the mass and the coupling such that there are no free parameters for the massive GN model typically. 
Those variational mass and coupling are to be simply used to evaluate 
the physical quantity to be optimized, giving a
physically relevant relation {\it e.g.} between the latter and the basic scale
$\Lambda_{\ms}$. Moreover it is completely
equivalent to optimizing {\em independently} simply with respect to the mass and 
coupling parameters.\\
Although our results are numerical and not a proof of convergence of the method, by 
adding the RG behaviour content of the theory in this most simple way we obtain successful approximations
of the exact mass gap below or at most at the percent level for arbitrary $N$ values, using only
the purely perturbative information. \\

One may wonder why the GN model behaves so well, in comparison
{\it e.g.} with the BEC model where the convergence appears slower~\cite{bec1,bec2,kastening2}, 
despite its super-renormalizable properties
implying that only the mass has non-trivial RG properties. 
Of course the GN model for large $N$ is a particularly simple 
theory, where the original perturbative coefficients do not exhibit the standard factorial divergences of most
field theory perturbative series. 
But for arbitrary $N$ the perturbative relation between the pole and 
Lagrangian mass, as well as RG properties,  
are very similar to those 
in more complicated theories, such as QCD typically. In fact,  even the large-order behaviour of the coefficients
of this perturbative series in the massive GN case have a similar, badly factorially divergent, 
(renormalon) behaviour~\cite{KRcancel}.
   In that respect the results obtained here with a fast numerical convergence at the percent level or less,
and controllable from understood RG behaviour, are to be considered 
very encouraging for similar analysis in more involved 
$D=4$ theories where $D=2$ exact S-Matrix and TBA results are inapplicable, and where the mass gap and 
other related non-perturbative quantities are not known and usually not expected to be 
derivable from first principles. This will be the object of future works.

\end{document}